\documentclass[aps,prl,twocolumn,superscriptaddress,showpacs]{revtex4-2}
\usepackage{graphicx,svg,amsbsy,amsmath,epsfig,float}
\usepackage{amssymb,latexsym,wasysym,pifont,mathrsfs,gensymb}
\usepackage{comment,verbatim,multirow,array,sidecap,color}
\usepackage{lineno,setspace,soul,cancel,enumerate,amsmath,mathtools}
\usepackage[normalem]{ulem}
\usepackage{bm,bbm}

\begin{document}
\title{Elastoplastic Modelling of Cyclic Shear Deformation of Amorphous Solids}
\author{Pushkar Khandare}
\author{Srikanth Sastry}
\email{sastry@jncasr.ac.in}
\affiliation{Jawaharlal Nehru Centre for Advanced Scientific Research, Jakkur Campus,  Bengaluru, 560064, India}


\begin{abstract}
We develop an energy-landscape based elasto-plastic model to understand the behaviour of amorphous solids under uniform and cyclic shear. Amorphous solids are modeled as being composed of mesoscopic sub-volumes, each of which may occupy states - termed mesostates -- drawn from a specified distribution. The energies of the mesostates under stress free conditions determine their stability range with respect to applied strain, and their plastic strain, at which they are stress free, forms an important additional property. Under applied global strain, mesostates that reach their stability limits transition to other permissible mesostates. Barring such transitions, which encompass plastic deformations that the solid may undergo, mesostates are treated as exhibiting linear elastic behavior, and the interactions between mesoscopic blocks are treated using the finite element method. The model reproduces known phenomena under uniform and cyclic shear, such as the brittle-to-ductile crossover with annealing and the Bauschinger effect for uniform shear, qualitative features of the yielding diagram under cyclic shear including the change in yielding behaviour with the degree of annealing, across a `threshold level', and dynamic phenomena such as the divergence of failure times on approach to the yield point and the non-monotonic evolution of the local yield rate. In addition to these results, we discuss the dependence of the observed behaviour on model choices, and open questions highlighted by our work. 

\end{abstract}
\maketitle

\textit{Introduction --- } Amorphous materials, which comprise of soft solids like gels, foams \citep{ysm_rmp_17} and hard solids like metallic glasses \citep{schuh07,bonfanti25} show a rich phenomenology when subject to applied stress or deformation. Subjected to uniform shear deformation, amorphous materials show solid-like elastic response which is punctuated by stress drops as deformation progresses, eventually leading to material failure \cite{argon79,falk98,lacks99,pollard_22,berthier25_review}. The yield process can be catastrophic, with a stress overshoot and subsequent localisation of strain, termed as shear banding \citep{divoux11}, or gradual and homogeneous without a stress overshoot \citep{lacks99}. The nature of yielding depends on preparation history \cite{falk05,ohern17,ozawa_18,lerner2021} (rapidly cooled glasses show ductile behavior, while slowly cooled glasses display a stress overshoot) and on the rate of deformation \cite{mukai02,berthier_03,berthier20,fielding20}. The yielding transition of glasses under cyclic shear \cite{priezjev2013,sastry13,regev_onset_2013,priezjev2013,kawasaki16,leishangthem17,parmar2019,berthier_prl_20,Das2020,bhaumik21,khirallah_21,sastry2021,mungan_sastry_21,parley_2022,liu_etal22,dheeraj_22,fielding24,maity_24,sarkar25,Suda2025} reveals a richer phenomenology. It was shown that the yielding transition is discontinuous, in the athermal quasistatic (AQS) limit, irrespective of the preparation history \cite{kawasaki16,leishangthem17,bhaumik21}, and that the time taken to reach the final state appears to diverge on approaching the yielding transition amplitude from either side \cite{sastry13,regev_onset_2013,kawasaki16,leishangthem17,dheeraj_22,fielding24,maity_24,Suda2025}. Particularly striking is the emergence of a threshold energy wherein samples with inherent structure (IS) energies above and below this threshold (termed \textit{poorly} and \textit{well} annealed respectively) show qualitatively different yielding behaviour; and that the threshold energy corresponds to the temperature where a dynamical cross-over is observed \cite{bhaumik21}. Failure under cyclic deformation occurs \textit{via} the formation of a shear band irrespective of the annealing level of the sample \cite{parmar2019,bhaumik21}, and samples show non-monotonic evolution of energies and local yield rate \textit{en route} to failure \cite{parmar2019,parley_2022,maity_24,sarkar25}. At low values of cyclic shear amplitude poorly annealed samples evolve towards absorbing states with lower energy, with all poorly annealed samples reaching a universal state termed the threshold state at a common yield amplitude. Beyond this amplitude the poorly annealed samples yield, with all samples tracing the same energy {\it vs.} strain amplitude curve, post-yield. Well annealed samples show negligible response to cyclic shear till a critical amplitude (greater than the yield amplitude of poorly annealed samples) is crossed, beyond which they follow the master curve that defines the universal yielded state that does not possess memory of the initial state. The deeper the annealing level below the threshold, higher is its critical amplitude.

The localized nature of plastic events in amorphous solids \cite{argon79,falk98} has prompted development of elastoplastic models (EPMs) (see \cite{epm_rmp18} for a comprehensive review) which regard an amorphous solid to be composed of mesoscopic blocks that are coupled elastically. When the local stress (equivalently, strain) exceeds the local yield value, a plastic rearrangement occurs whereby the mesoscopic block experiences a stress drop, and a transition into a new local state. This new local state is characterised by a new plastic strain. Although many EPMs have been investigated using a single local yield stress value, those EPMs (and corresponding mean field treatments) aiming to incorporate the heterogeneity of amorphous solids consider a a distribution both of the local yield stress values, and the plastic strain increments \cite{sollich97,agoritsas_relevance_2015,parley_2022,dheeraj_22,liu_etal22}. 

With the aim of capturing the annealing and rejuvenation effects observed under cyclic shear, here we develop an EPM based on an energy landscape picture that represents the state of each mesoscopic region with so-called mesostates \cite{sastry2021,mungan_sastry_prl_2019}. Each mesostate is a set of stable configurations that can be continuously transformed into each other \textit{via} changes in strain, within a stability range in strain values. Each un-strained mesostate may be viewed as a local energy minimum (or inherent structure) of the mesoscopic block, with a corresponding density of states (DOS) we specify. Reaching either limit (we consider a scalar strain variable here) of the stability range will lead to a discontinuous transition to a new mesostate. It has been observed \cite{bhaumik21} in simulations that the stability range of deeper lying mesostates was larger in simple model glasses (see \cite{Chatterjee2024} for variations associated with the fragility of the corresponding glass formers), which we incorporate in our model. To complete the prescription we need to specify how the plastic strain increment is chosen. A single site model with these ingredients \cite{sastry2021} captures several features of cyclic shear yielding, and has been the basis for further theoretical analysis of yielding and fatigue failure \cite{mungan_sastry_21,parley_2022,sarkar25}. In the present work, we construct an EPM with the properties of a mesoscopic blocks as described above, whose elastic interactions are treated employing the finite element method (FEM), following previous work \cite{nicolas2015,sandfeld15}. A popular alternative approach \cite{picard04,talamali12} uses the response of an infinite homogeneous elastic medium to a point plastic strain, which is then adapted to a finite periodic system, while others \cite{jagla07,cao_18,liu_etal22} account for the elastoplastic response by assuming a local disordered potential (with homogeneous elastic constants) with multiple minima. 

The FEM approach treats the plastic strain to involve an extended region, as opposed to a point, and easily permits incorporation of heterogeneous elastic moduli. The computational cost in the homogeneous case is comparable to applying the analytically derived kernel, as the FEM response needs to be calculated only once for a unit plastic strain and the response to any given plastic strain field is then found by linear superposition. Our work is closely related to a recent investigation on cyclic shear phenomena using energy landscape based approaches \cite{liu_etal22,dheeraj_22}. We systematically tune the model parameters, specifically the density of states and the plastic strain increment choice, and find a somewhat surprising sensitivity to specific choices. In addition to reproducing the phenomenology reported in earlier related works \cite{liu_etal22,dheeraj_22}, our results also show: i) the possibility of \textit{trenching}, depending on model paramters, where the shear bands remain pinned for indefinite number of shear cycles, ii) non-monotonic evolution of the energy and local yield rate in the yielded regime, and iii) divergence of number of cycles needed to form a shear band, which are consistent with simulation results. Further, our work highlights the presence of an {\it intermediate} regime around yielding as a new feature, that needs further investigations to elucidate. 

\begin{figure}
    \centering
    \includegraphics[width=0.8\linewidth]{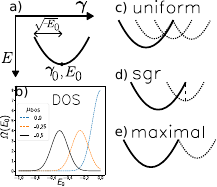}
    \caption{(a) A single mesostate with energy $E(\gamma) = E_0 + \frac{\mu}{2}(\gamma - \gamma_0)^2$ is stable over a finite range in strain, $\gamma \in (\gamma_0 - \sqrt{-E_0},\gamma_0 + \sqrt{-E_0})$. (b) Three choices of Gaussian density of states are shown, with the standard deviation fixed at $0.1$. (c,d,e) Plastic strain increment rules. For the results in the main text the model choices are: $\mu_{DOS}=-0.5$ and \textit{maximal} plastic increment.}
    \label{fig:model}
\end{figure}

\textit{Model --- }A scalar single site model was previously introduced \cite{sastry2021} which prescribes an energy landscape that is available to a single mesoscopic region undergoing shear deformation. Each region, hereafter termed mesoblock, shows linear elastic response around a stress-free plastic strain. The energy $E_{0i}$ at the plastic strain $\gamma_{0i}$, and a stability range around the plastic strain state characterise the full mechanical behaviour of a mesoblock $i$. The elastic energy of a mesoblock thus reads, $E_i = E_{0i} + \frac{\mu}{2}(\gamma_i - \gamma_{0i})^2$, and we assume that deeper lying mesostates have a larger stability range, specifically, a mesoblock $i$ is stable if $\gamma_i \in [\gamma_{0i} - \sqrt{-E_{0i}},\gamma_{0i} + \sqrt{-E_{0i}}]$. Once the current mesoblock becomes unstable the $E_0$ value of the new mesostate is taken from a density of states of energy minima $\Omega(E_{0})$. $E_0$ values are chosen to lie in the range $(-1,0)$. We choose $\Omega(E_0)$ to be a Gaussian with mean $-0.5$ and standard deviation $0.1$. Transition to another mesostate is made with the constraint that the local elastic energy be lower at that local strain value, which implements the athermal protocol. The plastic strain value of the new mesostate can be chosen \textit{via} any of the three protocols sketched in Fig. \ref{fig:model}: uniform, soft glassy rheology (SGR) and maximal. The uniform choice consists of choosing with uniform probability a $\gamma_0$ such that the new mesostate is stable at that strain value, while the SGR choice sets the new plastic strain to be the current local $\gamma_i$. It was shown \cite{sastry2021} that the single-site yielding diagram is robust to these two choices, giving qualitatively similar phenomenology. Here we use the \textit{maximal} protocol employed in \cite{liu_etal22} where the new plastic strain is maximally away from the current plastic strain. 
The model we investigate consists of $L\times L$ mesoblocks coupled elastically, with the elastic couplings treated using the finite element method. We prepare samples of different degrees of annealing by first deriving the analytical occupation probability $P(E_0,T_p)$ of mesostate minima given a parent temperature $T_p$, and use a zero-mean Gaussian distribution of initial plastic strains that gets narrower as $T_p$ decreases. Details of the implementation and additional results, including the dependence on choice of model features, are given in the Supplemental Material (SM) \cite{supp}.  

\begin{figure}[h!]
    \centering
    \includegraphics[width=1.0\linewidth]{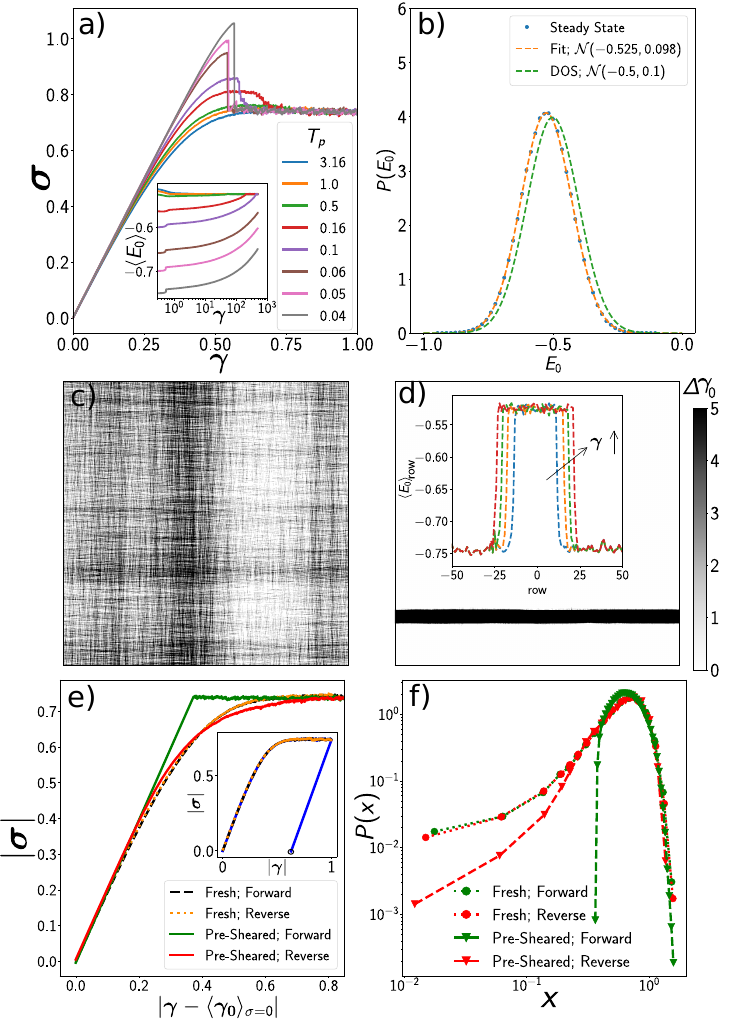}
    \caption{(a) The stress $\sigma$ is plotted for strain-controlled uniform deformation for various annealing levels as indicated by the parent temperature $T_p$. (Inset) Energy $E_0$ evolution is plotted. (b) Distribution of local $E_0$ in the steady state is plotted with blue points and dashed orange line denotes a Gaussian distribution with mean $-0.525$ and standard deviation $0.098$. The density of states is plotted for comparison. (c,d) The change in the plastic strain field between configurations at $\gamma = 2.0$ and $\gamma = 4.0$ is plotted for (c) poorly annealed ($T_p = 3.16$) and (d) well-annealed sample ($T_p = 0.04$. The plastic activity is diffuse for PA samples while it is strongly localized inside a band for WA samples. (inset) Row-averaged $E_0$ profile is plotted with the innermost blue curve at $\gamma = 4.0$ till the outermost red curve at $\gamma = 20.0$ in steps of $\gamma = 4.0$. (e) Bauschinger Effect: The forward and reverse strain response is plotted for a fresh sample (dashed and dotted lines respectively) and a pre-sheared sample (solid lines). Significant anisotropy in forward and reverse response can be observed for the pre-sheared sample. The inset shows the initial loading-unloading curves. (f) The distribution of local distances to strain thresholds is plotted in either direction for the pre-sheared and fresh sample. System size is $L=512$.}
    \label{fig:uni_composite}
\end{figure}

\textit{Results:Uniform Shear --- } Uniform shear deformation (see Fig. \ref{fig:uni_composite}a) reveals a ductile to brittle crossover as the parent temperature ($T_p$) is lowered. At large values of the strain ($\gamma$) all samples reach a common steady state stress value, $\sigma_{SS} = 0.737$, while the corresponding $\langle E_0 \rangle$ values reach a common value of $-0.525$. Note that this is lower than the peak of the DOS since the stability range increases as $E_0$ decreases which induces a bias to the negative side of the peak of the DOS. Poorly annealed samples (with higher $T_p$) exhibit ductile behaviour and reach this common $\langle E_0 \rangle$ value much faster than the brittle samples since poorly annealed samples display system-wide homogeneous plastic activity. The steady state solid resembles a high temperature system. This is evidenced from the invariant distribution (of $E_0$) that is reached in the flow stress regime, see Fig. \ref{fig:uni_composite}b. The distribution mimics that of a system corresponding to $T_p = 0.4$ \cite{supp}. Well annealed samples, on the other hand, fail with a stress overshoot followed by a stress drop, with the plastic activity being confined to a banded region which is termed as a \textit{shear band}. This band grows with increasing strain (inset of Fig. \ref{fig:uni_composite}d), with a square root growth law \cite{supp}; this slow growth is captured in the $E_0$ evolution as well. Once this shear band has grown to the linear extent of the system we expect total erasure of initial conditions.

The Baushchinger effect is an example of mechanical memory where we observe a softening of the response in the direction opposite to that in which the sample was previously deformed \cite{karmakar_proca_bauschinger,memory_rmp,anael20}. When we probe the response of a freshly prepared sample to shear strain in either direction, the response is found to be isotropic, see Fig. \ref{fig:uni_composite}e. However when the sample is sheared in the  positive direction and the strain is reversed till the stress is zero (see inset, \ref{fig:uni_composite}e), there is a change -- the response in the direction opposite to the initial loading is softer, see Fig. \ref{fig:uni_composite}e. In Fig. \ref{fig:uni_composite}f we plot the distribution of distance to the respective stability thresholds ($\gamma^+$ for forward, and $\gamma^-$ for reverse loading) for the fresh and the pre-sheared samples. The distributions in either direction are equivalent for the fresh sample while the distribution of forward distances are severely depleted in the low $x$ regime for the pre-sheared sample. This is due to a progressive removal of low $x$ sites during the preparation of pre-sheared sample as reported previously in \cite{talamali12}.

\begin{figure}
    \centering
    \includegraphics[width = 1.0\linewidth]{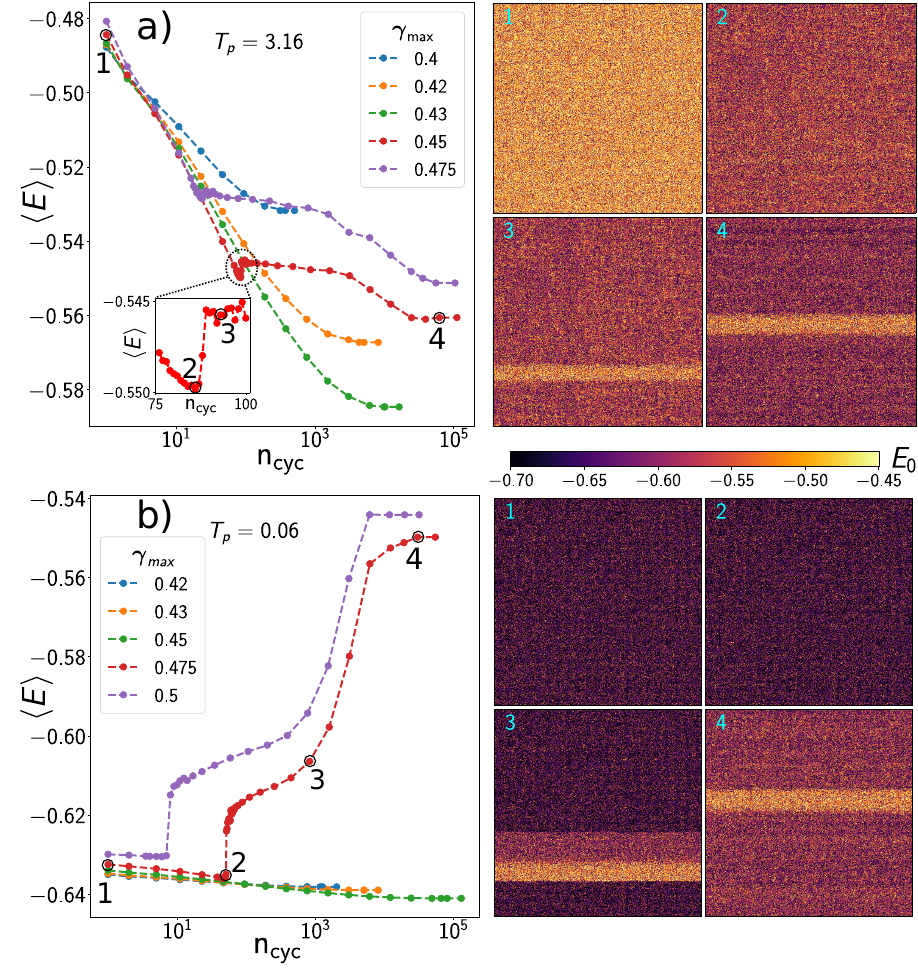}
    \caption{(a) Per site elastic energy ($E$) evolution is plotted for (a) poorly annealed sample ($T_p = 3.16$) and (b) well annealed sample ($T_p = 0.06$) for a range of $\gamma_{max}$ values below and above the yield point. Color-maps of $E_0$ field for configurations taken at points numbered $1-4$ are shown to the right of the respective panel.}
    \label{fig:pa_wa_composite}
\end{figure}

\textit{Cyclic Shear --- } In Fig. \ref{fig:pa_wa_composite}a we plot the evolution of the stroboscopic mesostate energy $E$ (at the end of each cycle, at zero strain) for a poorly annealed sample ($T_p = 3.16$). For driving amplitude $\gamma_{max} \leq \gamma_{max}^{yield}$ ($= 0.43$), we see that the energy drops with number of cycles and reaches an absorbing state. For higher values of  $\gamma_{max}$ we observe initial annealing not unlike the evolution towards the absorbing case, followed by a sharp up-jump when a shear band forms and the subsequent evolution involves motion of this shear band which anneals the system further, which explains the dip in energy seen for some cases at large number of cycles. $E_0$ maps are plotted on the right hand side, corresponding to configurations taken from points $1-4$ indicated in Fig. \ref{fig:pa_wa_composite}a. The higher-energy banded structure is the shear band. In Fig. \ref{fig:pa_wa_composite}b we repeat the same analysis for a well annealed case ($T_p = 0.06$). For driving amplitudes at and below the critical value of $\gamma_{max}^{yield} = 0.475$ an absorbing state is reached with negligible change in the mean energy. At higher $\gamma_{max}$ initial minimal annealing is followed by a sharp jump that signals the formation of a shear band. Further evolution denotes the shear band movement, moving ballistically \cite{liu_etal22} till all the sites are visited and then diffusing indefinitely in the solid.

These results are largely consistent with behaviour observed in simulations \cite{parmar2019,bhaumik21,maity_24}, except for the secondary annealing observed. This appears to be a feature of EPMs (seen also in \cite{liu_etal22}), but whether it faithfully captures the behaviour of glasses is unclear at present. In similar vein, we find that the yield point is not sharp; a \textit{coexistence} region exists in the vicinity of the yield point where the probability of evolving to an absorbing state drops from $1.0$ to $0.0$ over a finite range (of typical size $0.02$) of driving amplitudes (data shown in \cite{supp}). Interestingly, it is difficult to assess, with the available data, whether the width will vanish in the limit $L \rightarrow \infty$ (see \cite{supp}), and the significance of this co-existence region remains to be understood by future work. 

\begin{figure}[h!]
    \centering
    \includegraphics[width = 1.0\linewidth]{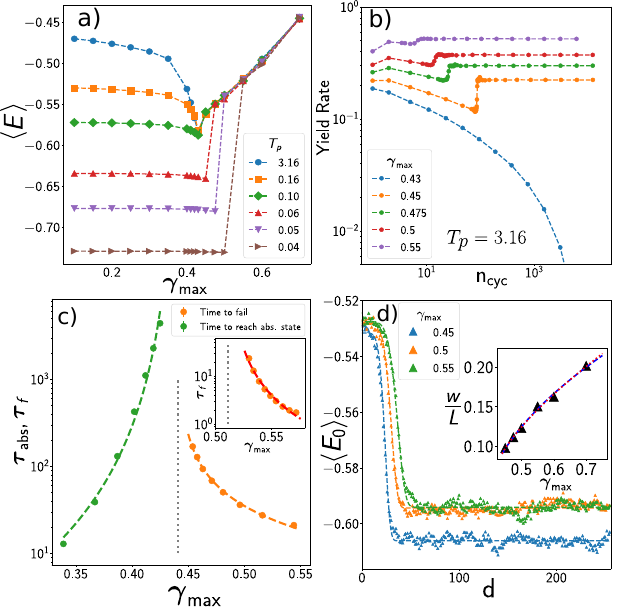}
    \caption{(a) Steady state energies are plotted at various shear amplitudes for various degrees of annealing (b) Evolution of number of plastic events per site per cycle - the yield rate - is plotted for a poorly annealed sample ($T_p = 3.16$) above and at the yield point $\gamma_{max}^{yield} = 0.43$. (c) Number of cycles to reach an absorbing state ($\tau_{abs}$) and to fail ($\tau_f$) are plotted for a poorly annealed sample. Dashed lines are power-law fits with exponents $-3.2$ ($\tau_{abs}$, green) and $-1.02$ ($\tau_f$, orange); vertical line denotes the yield amplitude in the main panel and the inset. (Inset) $\tau_f$ is plotted for a well-annealed sample. Dashed line shows a power-law fit with exponent $-2.2$. (d) Energy profiles averaged along the direction of the shear band are shown (centered by hand at zero, only the right half is plotted) for a poorly annealed sample ($T_p = 3.16$). Dashed lines are fits to a flat top Gaussian profile $c_1 + c_2 e^{-(x/c_3)^6}$. (Inset) The shear band width is plotted as function of the driving amplitude. Dotted  red line is a power law fit $w/L = w_0/L +  A(\gamma_{max} - \gamma_{max}^{yield})^{0.73}$. Blue dashed line denotes the fit, $w/L = A(\gamma_{max} - \gamma_y)^{1/2}$, where $\gamma_y$ is a free parameter, with the fit value being $\gamma_y = 0.373$.}
    \label{fig:cyclic_composite}
\end{figure}

We define the mid-point of the \textit{coexistence} region to be the yield point and report time-divergence data outside the \textit{coexistence} region. In Fig. \ref{fig:cyclic_composite}a we plot the steady state (or absorbing state) stroboscopic energies as a function of driving amplitude for various degrees of annealing. For loading amplitudes till $\gamma_{max}^{yield} = 0.43$, all samples reach a stable state where we see complete cessation of plastic activity. Beyond $\gamma_{max}^{yield}$ poorly annealed samples exhibit a shear band, and the system reaches a steady (rather than a stable) state. For well annealed samples, the transition occurs at progressively larger $\gamma_{max}$ values. Above the yield point, all initial conditions trace the same curve implying complete erasure of memory of initial conditions. The initial conditions are erased by way of motion of the shear band which moves throughout the sample. As mentioned, this feature may or may not reflect the corresponding situation in molecular glasses. 

In Fig. \ref{fig:cyclic_composite}b we plot the number of yield events undergone per site per cycle (yield rate) for a poorly annealed sample. We see a clear non-monotonicity, a reduction in yield events which is abruptly followed by at upturn upon failure, where a steady yield rate is observed, stemming from plastic activity in the shear band. A satisfactory qualitative match can be observed with the analytically derived yield rate reported in \cite{parley_2022}. 

The average time to failure is plotted for a poorly annealed sample ($T_p = 3.16$) in Fig. \ref{fig:cyclic_composite}c (orange points), while the same for a well annealed sample ($T_p = 0.04)$ is plotted in the inset. $1000$ samples were considered for each driving amplitude and the system size was $L=128$. A power law behaviour is observed, $\tau_f \propto (\gamma_{max} - \gamma_{max}^{yield})^{-\beta}$, where $\beta = -1.02 \pm 0.02$ for the poorly annealed sample and $\beta = -2.2 \pm 0.1$ for the well-annealed case. Recent 3D particle simulation results \cite{maity_24} observe an exponent of $-2$ irrespective of annealing level while preliminary results for 2D glasses suggest an  exponent of $-1$ \cite{himangsu_pcom}. Mean field investigations \cite{parley_2022,sarkar25} present a failure time exponent of $-1$ for poorly annealed samples and $-1/2$ for well annealed samples. While the lack of agreement with mean field results may be attributable to the approximations involved in those calculations, the deviation from simulation results is puzzling and needs an explanation. The apparent dimensionality dependence may offer an explanation, albeit a surprising one, which will be pursued further in future work. The time to reach an absorbing state for driving amplitudes below the yield point is plotted as green curve in the Fig. \ref{fig:cyclic_composite}c. Power-law divergence is observed with exponent $-3.2 \pm 0.1$, close to what is reported in \cite{regev_onset_2013,kawasaki16,khirallah_21,dheeraj_22}. A log-divergence was observed instead in \cite{liu_etal22} which we speculate is the behaviour of the system in the coexistence region.

Contrary to the results and expectation expressed in \cite{parmar2019,liu_etal22} the present results show that the region outside the shear band does not have an average energy that is independent of the strain amplitude. The yielded state at $\gamma_{\textrm{max}}$ values close to the transition shows a shear band diffusing in a background that is \textit{lower} in energy as compared to that for higher $\gamma_{max}$ values, see Fig. \ref{fig:cyclic_composite}d. For larger values of $\gamma_{max}$, the average energy outside the shear band appears more constant, but higher than the threshold energy. The shear band width $w$ follows a power law growth as distance from yield amplitude increases, $w/L = w_0/L + A(\gamma_{max} - \gamma_{max}^{yield})^\eta$, see red dotted line in the inset of Fig. \ref{fig:cyclic_composite}d, with a non-zero width fraction of about $7.5\%$ at the yield point, and the exponent is $0.73$. Following \cite{jagla10,falk_18,liu_etal22} if we enforce than the shear band width should follow a square-root growth law, that is, $w/L = A(\gamma_{max} - \gamma_y)^{1/2}$ where $\gamma_y$ is a fit parameter, we find $\gamma_y = 0.373 < \gamma_{max}^{yield}$ (see blue dashed line in inset of Fig. \ref{fig:cyclic_composite}d; within the quality of the data, these fits are indistinguishable) which also implies that at the yielding transition the shear band has a finite width of around $8\%$.



\textit{Discussion --- } We have introduced an elasto-plastic model building on the single-site mesostate model introduced in \cite{sastry2021}. The model reproduces several aspects of yielding behaviour under uniform and cyclic deformation that have been reported in previous work. These include, for uniform shear, the brittle-to-ductile crossover in yielding behaviour with annealing and the Bauschinger effect. For cyclic shear, the qualitative features of the yielding diagram are reproduced. Poorly annealed glassed exhibit mechanical annealing, and   evolve towards a unique threshold energy as the strain amplitude $\gamma_{\textrm{max}}$ is increased towards a common yield value. Well annealed glasses show negligible annealing before yielding at yield strain amplitudes that depend on, and increase with, the degree of annealing. Failure, upon repeated cycles of shear, is accompanied, close to the yield point,  by a non-monotonic evolution of energies and yield rate, and failure times (or below the yield point, cycles to reach absorbing states) exhibit power law divergences at the yield point. In addition these, however, we observe several new features that merit further investigation. The yielded state exhibits further annealing due to the movement of the shear band, that has no analog in simulations of glasses, a feature that further appears to be associated with the region outside the shear bands exhibiting average energies that differ from the threshold energy for different amplitudes, at variance with suggestions from previous work \cite{parmar2019,liu_etal22}. While power law divergences are observed for failure times, the exponent values do not agree with previous simulations and calculations. Such lack of agreement also leads to the intriguing suggestion that the failure time exponent may be dependent on the spatial dimensionality, which needs to be verified and rationalized. Finally, our results indicate the presence of an intermediate regime around the yield point, in which only a fraction of the samples investigated undergo failure; the width of this regime does not show a convincing reduction to zero as the system size increases. Such a feature, if it persists, is novel, and therefore merits further study. Although many of the results discussed are generically observed, some features, such as the pinning, {\it vs.} movement, of the shear bands, depend on the choice of model features in a manner that needs to be understood better. Our work thus  demonstrates the need, and paves the way, for future work on designing EPMs that accurately capture the rich phenomenology of amorphous solids. Obvious extensions include a fully tensorial three dimensional version, and calibration of model parameters to capture the properties of realistic glass models \cite{castellanos_22}.

\textit{Acknowledgements: } We thank H. Bhaumik, S. Maity, M. Mungan, A. Rosso, D. Sarkar and P. Sollich for useful discussions and comments on the manuscript. S. S. acknowledges SERB(ANRF) (India) for support through the JC Bose Fellowship (JBR/2020/000015) SERB(ANRF), DST (India) and a grant under SUPRA (SPR/2021/000382).


\bibliographystyle{apsrev4-2}
\bibliography{bibliography}

\clearpage
\onecolumngrid
\setcounter{page}{1}
\setcounter{figure}{0}
\setcounter{section}{0}
\setcounter{subsection}{0}
\setcounter{equation}{0}

\renewcommand\thepage{S\arabic{page}} 
\renewcommand\thefigure{S\arabic{figure}} 
\renewcommand\theequation{S\arabic{equation}}
\renewcommand\thesection{S\arabic{section}}
\renewcommand\thesubsection{\thesection.\arabic{subsection}}  
\renewcommand{\st}[1]{}

\centerline{\large{\textbf{Supplemental Material}}}

\section{Elasto-Plastic Model Implementation using the Finite Element Method}


We consider a two dimensional collection of $n\times n$ mesoblocks that are elastically coupled. We solve the linear elasticity problem via the finite element method \cite{zienkiewicz,nicolas2015}. The mesh is assumed to be a fixed, regular, square grid, with each element being a $4-\textrm{noded}$ square. The mesoblock of the elasto-plastic model directly corresponds to an element of this mesh. The FEM relies on the assumption that the deformation field at any point inside the continuum is calculated by interpolating the displacements on the nodes of the element that contains the point. Let the displacement at point $(x,y)$ inside any element $e$ be $\bm{u} = [u_x(x,y),u_y(x,y)]^T$, and the displacement at nodes is denoted by $\bm{u}_a$

    \begin{equation}
        \bm{u} = \sum_{a\in nodes}N_a (x,y) \bm{u}_a
    \end{equation}
These interpolation functions $N_a (x,y)$ are called \textit{shape functions}, where $N_a(x_a,y_a) = \bm{1}$, and additionally $N_a(x_b,y_b) = \bm{0}$ $\forall b \neq a$. For a $4-\textrm{noded}$ square element the simplest shape functions are bi-linear. The explicit form of the bilinear shape functions is (nodes are numbered $1-4$ anti-clockwise starting from bottom left, with $a$ being the side-length) \cite{zienkiewicz},
\begin{eqnarray}
    N_1 = \frac{(a - x)(a -y)}{a^2}; \quad
    N_2 = \frac{x(a-y)}{a^2}; \quad
    N_3 = \frac{xy}{a^2}; \quad
    N_4 = \frac{(a-x)}{a^2}
\end{eqnarray}

Now, we can write the strain in the element in terms of the nodal displacements on that element as follows,
 
\begin{equation}
    \begin{pmatrix}
        \epsilon_{xx} \\
        \epsilon_{yy} \\
        \sqrt{2}\epsilon_{xy}
    \end{pmatrix}
    = 
    \begin{pmatrix}
        \frac{\partial}{\partial x} & 0 \\
        0 & \frac{\partial}{\partial y} \\
        \frac{1}{\sqrt{2}}\frac{\partial}{\partial y} & \frac{1}{\sqrt{2}}\frac{\partial}{\partial x}
    \end{pmatrix}
    \begin{pmatrix}
        u_x \\
        u_y
    \end{pmatrix}
\end{equation}

\begin{equation}
    \bm{\epsilon}
    = 
    \begin{pmatrix}
            \frac{\partial}{\partial x} & 0\\
            0 & \frac{\partial}{\partial y}\\
            \frac{1}{\sqrt{2}}\frac{\partial}{\partial y} & \frac{1}{\sqrt{2}}\frac{\partial}{\partial x}
    \end{pmatrix}
    \begin{pmatrix}
            N_1 & 0 & \hdots & N_4 & 0 \\
            0 & N_1 & \hdots & 0 & N_4
    \end{pmatrix}
    \begin{pmatrix}
            u_{x_1} \\
            u_{y_1} \\
            \vdots \\
            u_{x_4} \\
            u_{y_4}
    \end{pmatrix}
\end{equation}

\begin{equation}
    \bm{\epsilon}
    =
    \begin{pmatrix}
            N_{1,x} & 0 & \hdots & 0 \\
            0 & N_{1,y} & \hdots & N_{4,y} \\
            \frac{1}{\sqrt{2}}N_{1,y} & \frac{1}{\sqrt{2}}N_{1,x} & \hdots & \frac{1}{\sqrt{2}}N_{4,x}
    \end{pmatrix}
    \begin{pmatrix}
            u_{x_1} \\
            \vdots \\
            u_{y_4}
    \end{pmatrix} \equiv \bm{B}\bm{u}^{e}, 
\end{equation}
where $\bm{u}^{e}$ is the $8 \times 1$ vector of node displacements, and $\bm{B}$, termed the \textit{elemental shape function}, is the $3 \times 8$ matrix of derivatives of the $N$'s ($N_{1,x} = (y - a)/a^2$, {\it etc.}). 
The elemental constitutive law gives us the stress-strain relationship. We assume isotropic and perfectly elastic behaviour, and the stress is given by the Generalised Hooke's Law
    \begin{equation}
        \sigma_{ij} = C_{ijkl}\epsilon_{kl}
    \end{equation}

In mechanical equilibrium the stress tensor is symmetric, and we use the Mandel notation to write it as a vector $\bm{\sigma}$, while the elasticity tensor has symmtries $C_{ijkl} = C_{jikl} = C_{ijlk} = C_{klij}$, and hence can be written as  $3 \times 3$ matrix (in $2$D) in the Mandel notation. For an isotropic solid in two dimensions, the elasticity tensor (in Mandel notation) reads,
\begin{equation}
    \bm{C}
    =
    \begin{pmatrix}
            K + \mu & K - \mu &  0 \\
            K - \mu & K + \mu & 0 \\
            0 & 0 & 2\mu
    \end{pmatrix}
\end{equation} where $K$ is the bulk modulus and $\mu$ is the shear modulus. We choose $K= 5\mu$ and $\mu=2$ ($K/\mu$ being in the higher end of typical values for bulk metallic glasses, and close to the ratio found for the two dimensional model glass \cite{castellanos_22}).

If we apply a small displacement $\delta\bm{u}$ at the nodes the work done by the stresses generated in the element over the elemental volume is $\int_{V} \delta\bm{\epsilon}^T\bm{\sigma} \textrm{dxdy}$, where $\delta\bm{\epsilon} = B\delta\bm{u}$. Assume that there are external forces acting on the nodes, denoted by $\bm{f}^{ext}$. The work done by these forces during to the virtual displacements would be $\delta\bm{u}^T\bm{f}^{ext}$. Since the system is in static mechanical equilibrium the total work done by the virtual displacements must vanish. Therefore we equate the external and internal work done and we get $\bm{f}^{ext} = \int_{V} \bm{B}^T\bm{\sigma} \textrm{dxdy}$. Therefore the force imposed \textit{by} the element \textit{on} the nodes of that element is opposite to the $\bm{f}^{ext}$. This gives
    \begin{equation}
        \bm{f} = -\int_{V} \bm{B}^T\bm{\sigma} \label{f2sigma} \textrm{dxdy}
    \end{equation}

We get,
\begin{eqnarray}
        \bm{f} &=& -\int_{V} \bm{B}^T\bm{\sigma} \textrm{dxdy} = -\int_{V} \bm{B}^T\bm{C}\bm{\epsilon} \textrm{dxdy}\\
        \bm{f}_{8\times 1} &=& -\left(\int_{V}[\bm{B}^T \bm{C}\bm{B}]_{8\times 8} \textrm{dxdy}\right) \bm{u}_{8\times 1}
    \end{eqnarray}
which is nothing but the elemental force-displacement relation. The integration is simple to perform in our case (more generally, Gaussian quadrature \cite{zienkiewicz} is used to perform the integration). The procedure followed eliminates numerical artefacts of hourglassing and checkerboarding without requiring the need for macro-elements (considering $2\times2$ elements as one element)\cite{zienkiewicz}.

We now construct the global force-displacement relation
    \begin{eqnarray}
        \bm{f}_{2N\times 1} &=& \bm{K}_{2N\times 2N}\bm{u}_{2N\times 1} \\
        \bm{u} &=& \bm{K}^{-1}\bm{f} \label{globalf2u}
    \end{eqnarray}
where $N$ is the total number of nodes and $\bm{K}$ is the global stiffness matrix, by taking into account the connectivity of each bulk node which is shared by four elements; this process is called \textit{assembly} \cite{zienkiewicz}.

%
    


Periodic boundary conditions compatible with simple shear deformation are enforced by constraining the degrees of freedom of matching nodes on opposing faces, specifically $u_x^{\mathrm{top}} - u_x^{\mathrm{bot}} = \gamma_{imp} L$ (where $\gamma_{imp}$ is the imposed shear deformation), and $u_y^{\mathrm{top}} = u_y^{\mathrm{bot}}$ alongwith $u_{\alpha}^{\mathrm{left}} = u_{\alpha}^{\mathrm{right}}$ with the bottom left node pinned to $(0,0)$ to remove global translational zero modes. These multi-node and single-node constraints can be written as $E\mathbf{u} = D$, where $\bm{u}$ is the full vector of $2N$ nodal displacements, and $N = L^2$ is the total number of nodes. There are $2(L-1)$ constraints relating the $x$ and $y$ components of pairs of nodes on top/bottom rows and $2(L-1)$ constraints relating the $x$ and $y$ components of pairs of nodes on the left/right columns. The bottom left corner is constrained, which implies that the rest of the three corners are constrained as well, specifically $u_x = u_y = 0$ for the bottom right corner and $u_x = \gamma_{imp} L$ and $u_y = 0$ for the top two corners. This gives a total of $(4(L-1) + 8)$ constraints. Therefore the size of $E$ is $(4L + 4) \times 2N$, with the $i$th row of $E$ being populated with zeroes except for the indices that select the of degree of freedom(s) that form the $i$th constraint. The column vector $D$ is populated by entries either $0$ or $\gamma_{imp} L$, through which the simple shear deformation of amplitude $\gamma_{imp}$ is enforced.

We must, in effect, solve Eq. \ref{globalf2u} subject to the constraints $E\mathbf{u} = D$. This is done {\it via} Lagrange multipliers $\bm{\lambda}$.

\begin{gather}
 \begin{pmatrix} \bm{K} & E^T \\ E & 0 \end{pmatrix}\begin{pmatrix} \bm{u} \\ \bm{\lambda} \end{pmatrix}
 = \begin{pmatrix} \bm{f} \\ D \end{pmatrix} 
\end{gather}

\begin{equation}
        \bm{u'} = \bm{K'}^{-1}\bm{f'} \label{lag_globalf2u}
\end{equation} 
which now becomes our modified global force-displacement relation.

The presence of a non-zero plastic strain changes the elastic energy function of a mesoblock. For concreteness let $e_1 = \epsilon_{xx} + \epsilon_{yy}$, $e_2 = \epsilon_{xx} - \epsilon_{yy}$ and $e_3 = \epsilon_{xy} + \epsilon_{yx} = 2\epsilon_{xy} = \gamma$, and let the plastic strain be only in mode $e_3$, with magnitude $\gamma_0$. The elastic energy of an element can then be written as,
\begin{equation}
E(e_1,e_2,e_3,\gamma_0) = E_{other}(e_1,e_2) +  \frac{1}{2}\mu (\gamma - \gamma_0)^2
\end{equation}
The force generated on any node due to this elastic energy can be found by taking the derivative of energy with respect to the deformation field on any node. Since the mesoblock energy is a function of elemental strains, it strictly depends only on the nodal displacements on the nodes of the element. We write out the complete expression for elemental energy with the strain component replaced with displacements, as we know $\mathbf{\epsilon} = \mathbf{B}\mathbf{u}$, where $\mathbf{B}$ is the elemental shape function and $\mathbf{u}$ is a vector of deformation field values on the $4$ nodes of that element.

Therefore, the force due to this elastic energy on the nodes of this element will be $F^{\alpha}_i = -\frac{\partial E}{\partial u_i^{\alpha}}$ where $\alpha = x,y$. Note that the force contribution due to elastic energy without any plastic strain has been taken care of in the local stiffness matrix. We only need the force contribution due to presence of a non-zero $\gamma_0$. Therefore, $F^{\alpha}_i = \Tilde{F}^{\alpha}_i + F^{\alpha}_{0i}$ where $F^{\alpha}_{0i}$ is a function of $\gamma_0$. 
\begin{eqnarray}
    F^{\alpha}_i &=& -\frac{\partial E}{\partial u_i^{\alpha}} \\
    F^{\alpha}_i &=& -\frac{\partial E_{\textrm{other}}}{\partial u_i^{\alpha}} - \mu(\gamma - \gamma_0)\frac{\partial \gamma}{\partial u^{\alpha}_i} \\
    F^{\alpha}_i &=& \Tilde{F}^{\alpha}_i + \mu\gamma_0\frac{\partial \gamma}{\partial u^{\alpha}_i} \\
    F^{\alpha}_{0i} &=& \mu\gamma_0\frac{\partial \gamma}{\partial u^{\alpha}_i}
    \label{eq:force_g0}
\end{eqnarray}

We know $\gamma$ in terms of $u^{\alpha}_i$ via the shape functions, specifically $\gamma = \sum_j \sqrt{2}B_{3j}u_j$ (note that the factor of $\sqrt{2}$ comes because we put it by hand as required by the Mandel notation \cite{nicolas2015}). We now have the force on the nodes that must be imposed to mimic the presence of a non-zero plastic strain. Whenever the global loading vector $\mathbf{f'}$ changes we use Eq. (\ref{lag_globalf2u}) to find the force-balanced displacement field $\mathbf{u}$ which then gives the elemental strains and stresses. We can understand this as follows: an element with a non-zero plastic strain will deform to that shape if it was cut out from the surrounding continuum. We have to externally deform it and put it back in the system. This region now pushes onto the elastic continuum till force balance is reached. Hence, the force in Eq. (\ref{eq:force_g0}) is the effect of having a non-zero plastic strain.

\begin{figure}
    \centering
    \includegraphics[width=0.7\linewidth]{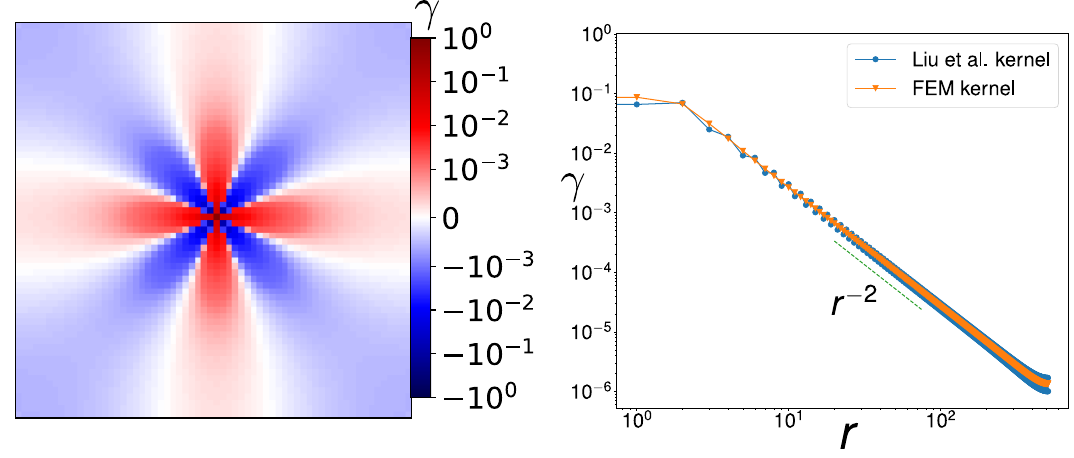}
    \caption{(Left) The shear strain field in response to a single mesoblock in the centre having a non-zero stress-free plastic strain $\gamma_0^{xy} = \gamma_0^{yx} = 1.0; \gamma_0^{xx,yy} = 0.0$. (Right) The analytically derived kernel from Liu \textit{et al.} and the FEM kernel show a $r^{-2}$ decay in the long field. The source value for the analytically derived kernel is $0.52$ and for the FEM derived kernel it is $0.47$.}
    \label{fig:gamma_resp_single_site}
\end{figure}

The response of the extended system when a single mesoblock is assigned a non-zero $\gamma_0$ value reproduces the familiar Eshelby quadrupolar structure as shown in the left panel of Fig. \ref{fig:gamma_resp_single_site}. We compare the response from the finite element method with the analytically derived response in Liu \textit{et al.} \cite{liu_etal22} in the right panel of the same figure, and find appreciable differences in the near-field region only, followed by a long-ranged $r^{-d}$ ($d$ is the dimension, here $d =2$) decay as expected from linear elasticity.  

\section{Numerical implementation of the Elasto-Plastic Model}

The elasto-plastic model algorithm goes as follows:
\begin{enumerate}
    \item Find the minimum strain increment (in the direction of the deformation) that takes exactly one site to its stability limit. Increment the globally imposed strain by this value.
    \item That site is made to undergo a transition, and the FEM solution is recalculated with the updated plastic strain field keeping the global strain constant. This can lead to some of the sites becoming unstable. 
    \item Unstable site(s), if any, are updated in parallel.
    \item If the recalculated local strain field keeps all mesoblocks stable then go to step $1$, else go to step $3$.
\end{enumerate}

Computational cost of step $2$ can be reduced significantly for the choice of a uniform elastic moduli throughout the sample. This keeps the global stiffness matrix a constant, and the response for given $\gamma_0$ field can be written as a linear combination of the response to a $\delta$ plastic strain field centered at each element. We calculate the $\delta$ response once and convolve this response with the given $\gamma_0$ field to find the strain field generated due to any arbitrary plastic strain field. Convolution, which is a non-local $O(N^2)$ operation becomes a local $O(N)$ frequency-wise multiplication in the Fourier space (here $N$ is the number of mesoblocks), and hence the Fourier transform of the $\gamma_0$ field is computed (using Fast Fourier Transform routines) whenever there are plastic events, and the local strain field is then calculated.

\section{Generating samples with different degrees of annealing}
\label{sec:gen_samp}

We generate initial distributions of states occupied by mesoblocks by estimating the probability of occupation of the mesostates at a given temperature $T_p$, assuming that all states are accessible according to their equilibrium probability. The density of states $\Omega(E_0)$ specifies the energy landscape sampled by any given mesostate. The form is assumed to be Gaussian as known from earlier work \cite{heuer_00,sastry_01} on supercooled liquids. At any given temperature the occupancy of these inherent structures (mesostates) is weighted by the basin contribution. We implicitly assume a harmonic approximation to the basins. Specifically, the form of the energy inside the basin is assumed to be of the form $E(\bm{q}) = E_{0} +\sum_{i=1}^{3N} k_iq_i^2$. The curvature along the strain direction is assumed to be given by the shear modulus $\mu$. We can now calculate the occupation probability for a mesostate to be found in a basin with inherent energy $E_0$. Schematically, it should be the bare density of states $\Omega(E_0)$ multiplied with the Boltzmann weight $e^{-\beta E_0}$ and the basin contribution. Thus,
\begin{equation}
    P(E_0,T) = \frac{\Omega(E_0) e^{-\beta E_0}C_{basin}(E_0,T)}{\int\Omega(E_0) e^{-\beta E_0}C_{basin}(E_0,T)dE_0}
\end{equation} where we assume $\Omega(E_0)$ is a Gaussian distribution with mean $\mu_{DOS}$ and standard deviation $\sigma$, that is, $\Omega(E_0) = A\mathrm{exp}[-(E_0 - \mu_{DOS})^2/2\sigma^2]$, with $A$ being the normalising constant.

In the harmonic approximation to the basin contribution one assumes that at low temperatures the system is found sampling the vicinity of the minimum, and hence experiences a quadratic potential. This implies that the basin contribution comes from the vibrational modes which depend on the $3N$ curvatures, and further these curvatures are assumed to be the same for all basins in the landscape. In the elasto-plastic model the basins along the strain direction are taken to be quadratic. We consider two ways to get the distribution of $E_0$ values for a given temperature $T$. In the first case, we assume that the basin contribution is the same for all basins. This will be labelled as $E_0$ independent basin contribution in what follows.

\begin{eqnarray}
    C^{indep.}_{basin}(T) &=& \int_{r^{3N} \in basin}e^{-\beta \Delta E} dr^{3N} = \Pi_{j = 1}^{j = 3N-3}\left(\frac{2\pi}{\beta k_i} \right)^{1/2} \\
    P^{indep.}(E_0,T) &=& \frac{Ae^{-(E_0 - \mu_{DOS})^2/2\sigma^2}e^{-\beta E_0}\Pi_{j = 1}^{j = 3N-3}\left(\frac{2\pi}{\beta k_i} \right)^{1/2}}{\Pi_{j = 1}^{j = 3N-3}\left(\frac{2\pi}{\beta k_i} \right)^{1/2}\int Ae^{-(E_0 - \mu_{DOS})^2/2\sigma^2} e^{-\beta E_0}dE_0} \\
    P^{indep.}(E_0,T) &\propto& \textrm{exp}\left(\frac{-(E_0 - [\mu_{DOS} - \beta\sigma^2])^2}{2\sigma^2}\right)
\end{eqnarray}

In the second case, we incorporate the basin-dependent finite range along the strain direction into the calculation, which minimally distinguishes the different basins by taking into account the differences in the stability range with respect to the application of strain. Specifically, the basin contribution is basin-independent along all directions but one - the strain direction.
\begin{eqnarray}
    C^{dep.}_{basin}(E_0,T) &=& \int_{\gamma^-(E_0)}^{\gamma^+(E_0)}e^{-\beta \Delta E} d\gamma \int_{r^{3N-1} \in basin}e^{-\beta \Delta E} dr^{3N-1} \\
    &=& \mathrm{erf}\left(\sqrt{\frac{-\beta \mu E_0}{2}}\right) \Pi_{j = 1}^{j = 3N-4}\left(\frac{2\pi}{\beta k_i} \right)^{1/2} \\
    P^{dep.}(E_0,T) &=& \frac{Ae^{-(E_0 - \mu_{DOS})^2/2\sigma^2}e^{-\beta E_0}\mathrm{erf}\left(\sqrt{\frac{-\beta \mu E_0}{2}}\right)\Pi_{j = 1}^{j = 3N-4}\left(\frac{2\pi}{\beta k_i} \right)^{1/2}}{\Pi_{j = 1}^{j = 3N-4}\left(\frac{2\pi}{\beta k_i} \right)^{1/2}\int Ae^{-(E_0 - \mu_{DOS})^2/2\sigma^2} e^{-\beta E_0} \mathrm{erf}\left(\sqrt{\frac{-\beta \mu E_0}{2}}\right) dE_0} \\
    P^{dep.}(E_0,T) &\propto& \textrm{exp}\left(\frac{-(E_0 - [\mu_{DOS} - \beta\sigma^2])^2}{2\sigma^2}\right)\mathrm{erf}\left(\sqrt{\frac{-\beta \mu E_0}{2}}\right)
\end{eqnarray}

\begin{figure}
    \centering
    \includegraphics[width = 0.75\textwidth]{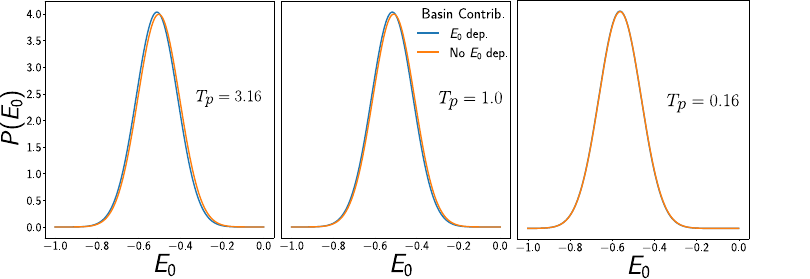}
    \caption{The occupation probabilities plotted for three temperatures $T_p = 3.16$ (left), $T_p = 1.0$ (middle), $T_p=0.16$ (right) using the $E_0$ dependent and $E_0$ independent basin contribution calculation.}
    \label{fig:compare_all_temp}
\end{figure}

We show the difference in resulting occupation probability distributions $P(E_0,T)$ from either method in Fig. \ref{fig:compare_all_temp}. We observe that the distributions, computed from either calculations, are slightly shifted with respect to each other (the mean is 2\% lower for the basin-independent calculation, while the standard deviation is lower by 1\%, at the highest temperature considered $T = 3.16$). Further, these differences become negligible as the temperatures are lowered. We prepare initial conditions using the $E_0$-dependent basin contribution calculation.

\begin{figure}
    \centering
    \includegraphics[width = 0.75\textwidth]{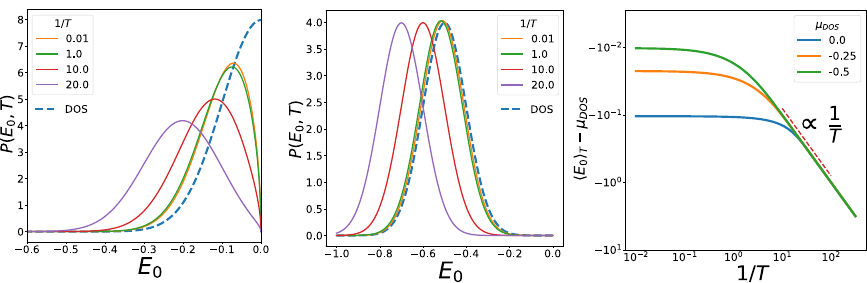}
    \caption{The occupation probabilities plotted for various temperatures for Gaussian DOS with:(Left) Mean $0.0$ and standard deviation $0.1$; (Middle) Mean $-0.5$ and standard deviation $0.1$. The DOS is truncated to lie between $(-1,0)$ in both cases. (Right) The dependence of mean $E_0 (T)$ calculated using the analytical occupation probability shows a clear $T^{-1}$ dependence. The calculation was done for three values of Gaussian DOS means, with the standard deviation held fixed at $0.1$.}
    \label{fig:thermal_dist_composite}
\end{figure}

Thus we see that, starting from a Gaussian density of states for $E_0$,  the occupancy at any temperature $T$ is the product of an error function that depends on temperature and a Gaussian distribution with the same standard deviation as that of parent density of states and a shifted mean with the shift being a function of the temperature, see Fig. \ref{fig:thermal_dist_composite}. At low temperatures (high $\beta$) the error function goes to unity. Hence, the mean $E_0$ at low temperatures goes linearly with $\beta$, specifically $\langle E_0 \rangle_T = \mu_{DOS} - \beta \sigma^2$.

We also need to prescribe the initial plastic strains that correspond to different degrees of annealing. We choose Gaussian distributed plastic strains with zero mean and a standard deviation that gets smaller as the parent temperature decreases. For the case where DOS is a Gaussian with mean $-0.5$ and standard deviation $0.1$, the standard deviation of initial plastic strains is chosen to be $[0.35,0.30,0.29,0.25,0.24,0.23]$ for $T_p \in [3.16,0.16,0.1,0.06,0.05,0.04]$ respectively.

\section{Determining the Yield Point}

\begin{figure}[h!]
    \centering
    \includegraphics[width = 0.75\textwidth]{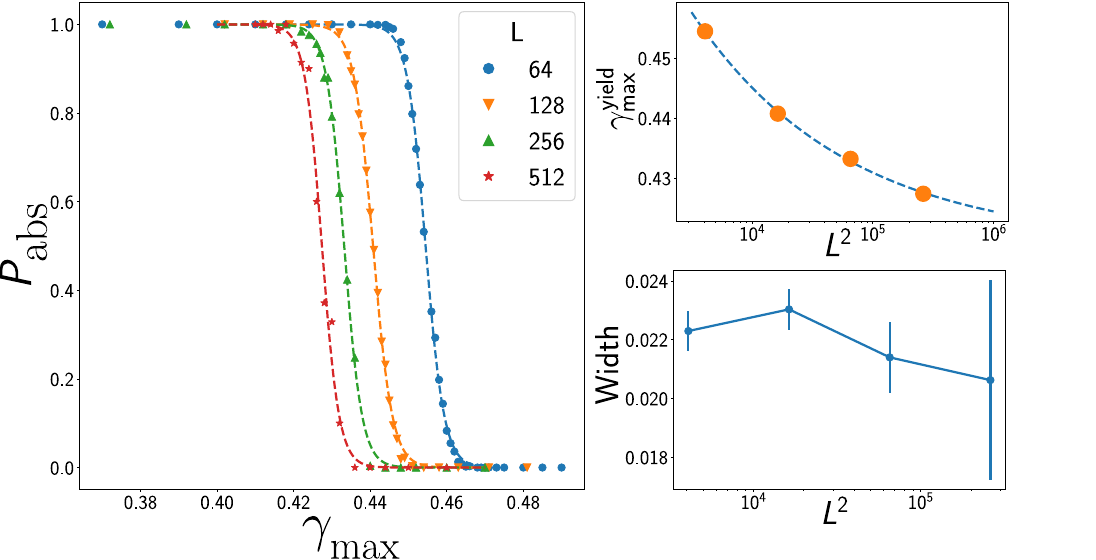}
    \caption{(Left) For a poorly annealed sample ($T_p = 3.16$) the probability of reaching an absorbing state is plotted as a function of the driving amplitude $\gamma_{max}$. (Right Top) The yield point $\gamma_{max}^{yield}$, identified as the value where $P_{abs} = 0.5$, shifts to lower values with increasing system size. (Right Bottom) The width of the \textit{coexistence} region is defined as the interval where $P_{abs}$ goes from $0.99$ to $0.01$. We do not see a clear decreasing trend of the width with increasing system size.}
    \label{fig:p_abs}
\end{figure}

To estimate the yield point we track the probability of a sample to fall into an absorbing state as the driving amplitude is varied. In Fig. \ref{fig:p_abs} we plot this probability for four system sizes $L=64,128,256,512$ for a poorly annealed sample corresponding to a parent temperature of $3.16$. We considered $1000$ samples for $L=64,128$, $250$ samples for $L=256$ and $70$ samples for $L=512$. For all the driving amplitudes considered we ensure that the samples were subject to as many cycles needed to either reach an absorbing state or fail via the formation of a shear band. For low driving amplitudes all samples reach an absorbing state and for high driving amplitudes all samples fail. We observe a \textit{coexistence} region where some samples fail while some samples reach an absorbing state. We use the logistic curve $(1 + e^{-k(x-x_0)})^{-1}$ as the fit function to describe the data, where $P_{abs} = 0.5$ at $x_0$ and $k$ controls the steepness. With the current data we cannot conclude if this coexistence region gets smaller with increasing system size. The widths of the driving amplitude region where the probability drops from $0.99$ to $0.01$ are $(0.022, 0.023, 0.021, 0.020)$ as the system size increases. We assign the yield point to be the mid-point of this logistic curve. This point systematically shifts to lower values - $(0.454,0.441,0.433,0.427)$ - as the system size increases and is found to fit well by $\gamma_{max}^{yield}(L) = \gamma_{max}^{yield}(L\rightarrow\infty) + aL^{-1/\nu}$, where we find $\gamma_{max}^{yield}(L\rightarrow\infty) = 0.419 \pm 0.003$, and $\nu = 2.9 \pm 0.4$, $a = 0.6 \pm 0.2$.

\section{Shear band growth in uniform shear}

\begin{figure}[h!]
    \centering
    \includegraphics[width=0.8\textwidth]{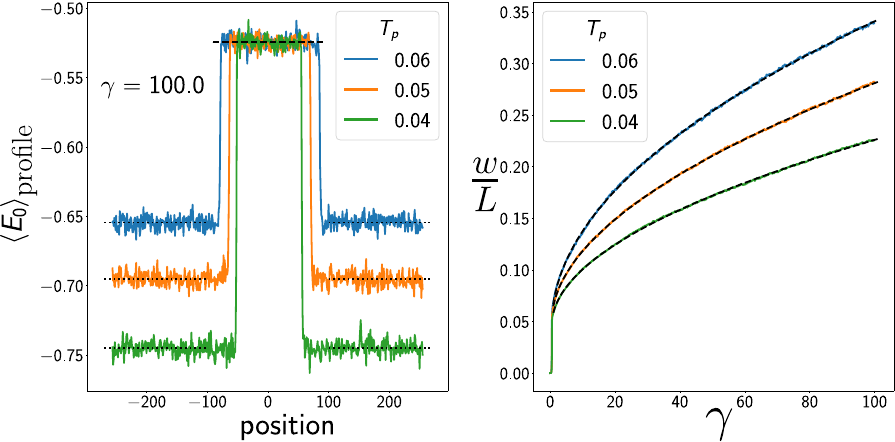}
    \caption{(Left) $E_0$ profiles averaged along the shear-band direction are plotted at $\gamma = 100.0$. Energy in the band (dashed black line) does not depend on parent temperature. Dotted black lines represent the average background energy for the three parent temperatures. (Right) The shear band fraction - extracted from $E_0$ evolution, energy in band and energy in background regions - is plotted as a function of $\gamma$. Fit lines (shown in black) correspond to $w_0 + c(\gamma - \gamma_y)^{\eta}$.}
    \label{fig:uni_sb}
\end{figure}

In this section we study the growth of the shear band that forms after the macroscopic stress drop in well-annealed samples when subject to uniform loading. In the left panel of Fig. \ref{fig:uni_sb} we plot the $E_0$ profile averaged along the direction of the shear band for samples at $T_p = 0.06,0.05,0.04$. The configurations are at $\gamma = 100.0$. We can see that the average $E_0$ value in the band is $-0.525$ irrespective of the parent temperature. The background region is populated with initial sites since no plastic events have occured there yet. Thus, the average energy of the background depends on the parent temperature. Further, the inset in Fig. 2d (in main text) shows that as deformation progresses the band widens. This allows us to write the average energy $E_0$ of the yielded solid as a function of $\gamma$ as follows. Let $w/L$ be the fraction of the solid that comprises the shear band, then,
\begin{eqnarray}
    \langle E_0 \rangle &=& E_{bg} (1 - \frac{w}{L}) + E_{b}\frac{w}{L} = E_{bg} + (E_{b} - E_{bg})\frac{w}{L} \\
    \frac{w}{L} &=& \frac{\langle E_0 \rangle - E_{bg}}{E_b - E_{bg}}
\end{eqnarray}
 where $E_b$ is the average energy in the band, and $E_{bg}$ is the average energy in the background region. Thus, we see that, $\langle E_0 \rangle$ is directly proportional to the fraction of the shear band and that we can extract the shear band fraction from the $E_0$ evolution data. In the right panel of Fig. \ref{fig:uni_sb} we plot the evolution of fraction of shear band (extracted from $E_0$) for three levels of annealing and fit the data to $w_0 + c(\gamma - \gamma_y)^{\eta}$ where $\gamma_y$ is the strain value where shear band formation occurs in the solid. We find $\eta = 0.5 \pm 0.02$ which matches well with the theoretical prediction in \cite{jagla10}.
 
\section{Trenching}
\begin{figure}[h]
    \centering
    \includegraphics[width=0.75\textwidth]{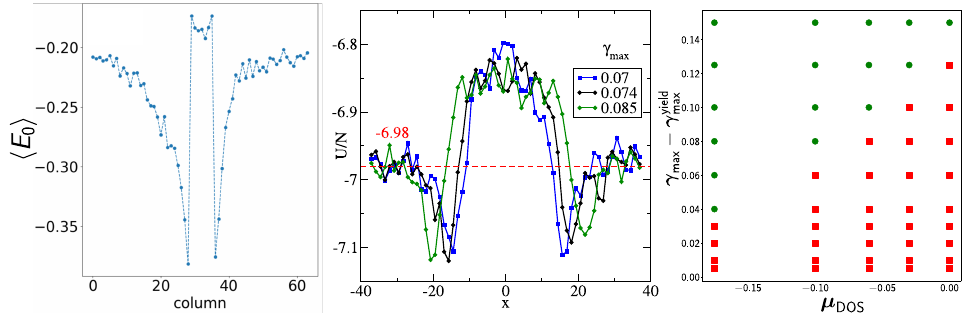}
    \caption{(Left) Trenching seen in EPM simulations for the half-normal density of states. (Middle) Trenching seen in SLLOD-MD simulations. The threshold energy is marked in red \cite{maity_pcom}. (Right) For a given DOS mean ($\mu_{DOS}$), with fixed standard deviation $0.05$, a range of driving amplitudes above yield are considered. The cases where shear band remain pinned is marked in filled red squares, while filled green circles denote a moving shear band. Data for $L=64$. The plastic increment choice was \textit{uniform}.}
    \label{fig:trench}
\end{figure}

An unexpected feature uncovered from our investigations is that of \textit{trenching}. This refers to the case where the shear band remains pinned throughout the duration of cyclic shear, after apparent failure. Deep lying sites develop on either side of the shear band, while the shear band itself is composed of an invariant set of shallow-lying states. The trenching effect bears a striking resemblance to the energy profiles extracted from molecular dynamics simulations where we see that the shear band indeed remains pinned, with well developed trenches on either side, see left and middle panels of Fig. \ref{fig:trench}. This pinning in the atomistic case remains even for driving amplitudes where the shear band width is approaching the linear system size. This resemblance, though interesting, is misleading. The erasure of initial conditions in the elasto-plastic model is only possible via motion of the shear band. It is unclear how a universal yielded state is reached in atomistic simulations given the fact that the shear band remains static, after a transient. Further, as will be discussed in the follwing section, the presence of trenching in EPM simulations leads to an anomalous yielding diagram. For the half-normal density of states we see trenching over a maximal range of driving amplitudes. As we shift the peak of the density of states away from zero, which in turn reduces the probability weight at $E_0 = 0$, we see that the range of driving amplitudes over which trenching is observed diminishes, as shown in rightmost panel of Fig. \ref{fig:trench}. Although the significance and origin of the phenomenon of trenching in real glasses is unclear at present and needs to be further explored, the results discussed here highlight that choices of model parameters can have a strong influence on the observed behaviour in the case of cyclic shear, whereas the emergent qualitative properties are much more robust for uniform shear. 

\section{Effect of DOS and plastic strain increment on the yielding transition under cyclic shear}

\begin{figure}[h!]
    \centering
    \includegraphics[width=0.9\textwidth]{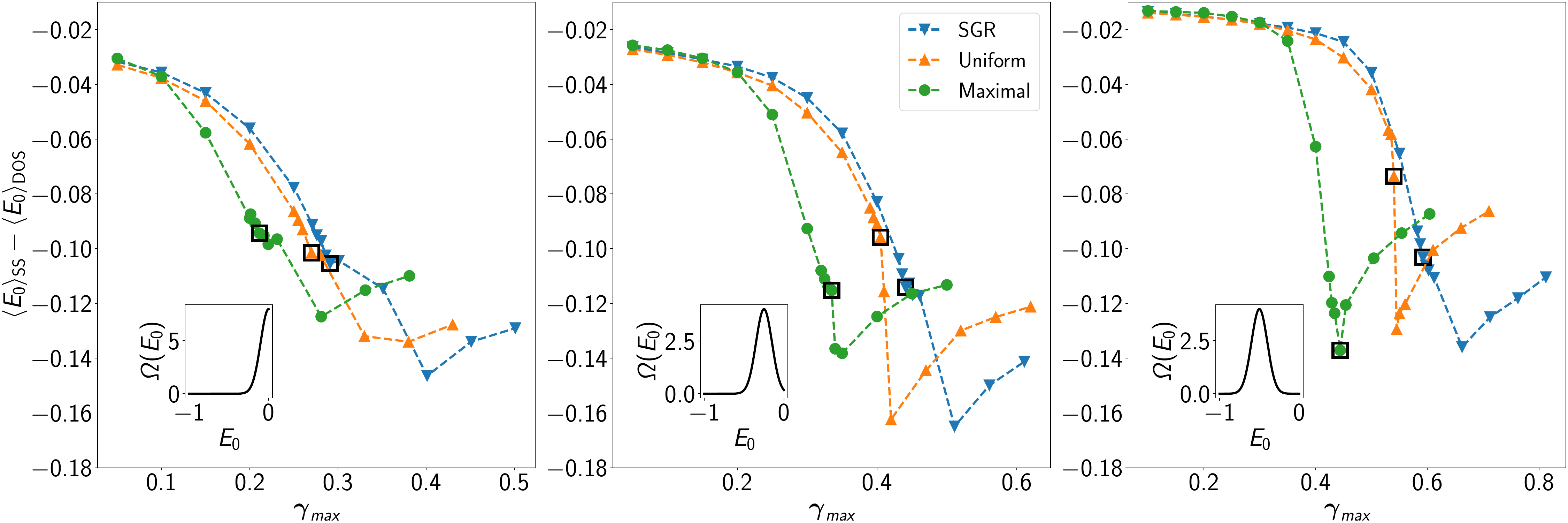}
    \caption{Steady state energies of a poorly annealed sample for DOS (plotted in the inset) with standard deviation $0.1$ and mean $0.0$ (left), $-0.25$ (middle) and $-0.5$(right). $\langle E_0 \rangle_{SS}$ denotes the steady state value while $\langle E_0 \rangle_{DOS}$ is the mean value of $E_0$ with respect the density of states $\Omega(E_0)$. Open square symbol (in black) marks the largest $\gamma_{max}$ value where an absorbing state was reached, that is, it denotes the threshold state. Different curves in each figure correspond to the three plastic increment choices considered.}
    \label{fig:yd_sweep}
\end{figure}

For three choices of the DOS ($\mu_{DOS} = 0.0,-0.25,-0.5$, with a fixed standard deviation of $0.1$) and three choices of plastic strain increment rule (SGR, uniform and maximal) per DOS, a corresponding poorly annealed sample was prepared and subjected to cyclic shear at various driving amplitudes with particular focus around the yield point. Our observations, summarized in Fig. \ref{fig:yd_sweep} (also see \cite{pushkar_thesis}), show that not all combinations of $\mu_{DOS}$ and plastic strain increment reproduce the expected yielding diagram. The probability weight at $E_0 = 0$ (where the stability range goes to zero) is maximum for $\mu_{DOS} = 0.0$ and appreciable for $\mu_{DOS} = -0.25$, while it is minuscule for $\mu=-0.5$ (see Fig. 1b in the main text). For the first two choices we do not recover the correct yielding behaviour for any of the plastic increment choices considered. For the case of $\mu_{DOS} = -0.5$ only the \textit{maximal} plastic increment choice reproduces the expected yielding diagram. Note that all choices show the yielding transition but only one among nine combinations considered reproduces the expected yielding diagram.

To reiterate, the expected yielding transition phenomenology (based on AQS simulations of atomic glasses) is that the system reaches maximal annealing - the threshold state - \textit{at} the yield point, not \textit{after}, with the yielded state having steady state energies strictly above the threshold energy. As the driving amplitude is increased beyond yield point the steady state energies are expected to increase monotonically. 

For $\mu_{DOS} = 0.0$, see left panel in Fig. \ref{fig:yd_sweep}, which is the half-normal case considered in \cite{sastry2021} where the single-site results reproduce the expected yielding diagram qualitatively, in the full model we see that even though the steady state energies of the just-yielded state lie above the threshold state, as the driving amplitude is increased the steady state energy \textit{decreases}. The yielded state in this case is comprised of a shear band which is \textit{pinned}. Such a pinned shear band is straddled on both sides by a narrow region of very stable mesostates, a phenomenon we term trenching (see preceding section). As the driving amplitude is increased the size of the pinned shear band increases, and at a particular value of the driving amplitude the shear band becomes \textit{mobile} (see the right panel of Fig. \ref{fig:trench}). The motion of this shear band further \textit{anneals} the system leading to a lower steady state energy than that attained for lower post-yield amplitudes where the band was pinned. This scenario is not remedied by any of the plastic increment choices.

For $\mu_{DOS} = -0.25$, see middle panel in Fig. \ref{fig:yd_sweep}, the just-yielded state is \textit{lower} in energy than the threshold state. In this case the shear band is not pinned, but the motion of the shear band anneals the system to an extent that the steady state energy lies significantly below the threshold energy. This implies that the threshold state reached was not optimal. This scenario is not remedied by any of the plastic increment choices. Qualitatively similar picture holds for $\mu_{DOS} = -0.5$ (right panel of Fig. \ref{fig:yd_sweep}) for the uniform and SGR plastic increment choice.

For $\mu_{DOS} = -0.5$, with the maximal choice of plastic strain increment, we observe the correct yielding diagram where the yielded branch lies above the threshold energy and increases monotonically with driving amplitude in the yielded region, and that the threshold energy reached is most negative when compared to uniform and SGR choices. Thus, a) the plastic strain increment rule must be \textit{maximal}, and b) the Gaussian density of states \textit{must} be shifted enough to have a vanishing weight at the zero-stability region. If we violate  the first requirement but satisfy the second we observe that the yielded state is lower than the threshold energy and that the band does not remain pinned. If we meet the first requirement and violate the second we start seeing pinned bands in a small region of driving amplitudes beyond yield, beyond which the yielded steady state energy is lower than the threshold energy. 

Further work is needed to elucidate why maximal plastic increment choice leads to maximal annealing, and why the presence of probability weight around $E_0 = 0$ leads to pinned bands. Once again, these result show that the observed behaviour in the case of cyclic shear yielding are not robuts with respect to model choices. The reasons, though partially apparent, need to be understood better in order to widely employ EPMs of the kind we present in this work. 




\section{Fatigue Failure}

\begin{figure}[h]
    \centering
    \includegraphics[width = 0.75\textwidth]{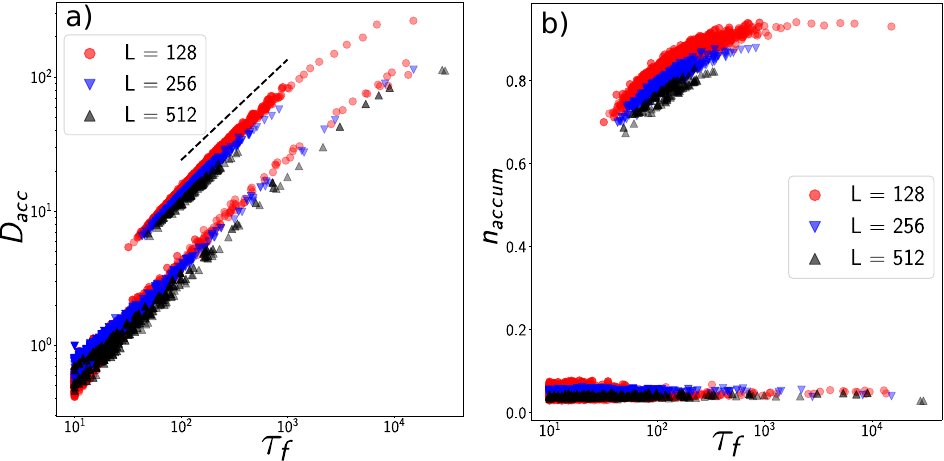}
    \caption{(a) Accumulated damage ($D_{acc}$) is plotted with respect to the failure time ($\tau_f$). The upper set of points are for a poorly annealed sample, while the lower set are for a well annealed sample. A power-law trend can be observed, the dashed line indicates an exponent of $0.75$. (b) Sites which underwent a plastic event are labelled as $1$, else $0$. The fraction of sites that are labelled $1$ by the time failure occurs is termed as $n_{accum}$. The upper set of points correspond to poorly annealed samples, while the lower set of points correspond to well annealed samples.}
    \label{fig:damage}
\end{figure}

Fatigue failure refers to the phenomenon of failure that occurs after repeated cycles of loading. In a recent study on fatigue failure in atomistic systems \cite{maity_24} the authors develop and use the notion of mobile particles and damage (defined as the stress-strain loop area) accumulated till failure initiation to uncover the process of fatigue failure. They found that the accumulated damage follows a power-law scaling with the failure initiation time, $D_{acc} \propto \tau_{fi}^{0.8}$ and when the failure time (mid-point of the transformation curves) is considered, $D_{acc} \propto \tau_{f}^{0.75}$ was found to describe the data well. We have done a similar analysis on the model data, computing failure times, and observe a similar phenomenology. We find that an exponent of $0.75$ describes the model data well, see Fig. \ref{fig:damage}a. Further, the authors devised a procedure to label the particle(s) which are involved in plastic rearrangements, which were termed as \textit{mobile} particles. These \textit{mobile} particles, when accumulated till failure, lead to a very interesting observation: failure occurs when the fraction of accumulated mobile particles reaches some constant value, irrespective of the amplitude of strain and sample to sample variations. This constant value is different for poorly annealed and well annealed samples. In the present model, it is straightforward to locate the sites which underwent plastic reorganization \textit{en route} to failure. A similar analysis confirms the observation of a constant fraction $n_{accum}$ at failure for well-annealed samples, but the fraction depends on failure time for poorly annealed samples, see Fig. \ref{fig:damage}b. The latter feature may be an outcome of an over-counting of mobile sites, similar to the ambiguity encountered in the simulations reported in \cite{maity_24}. Thus, although the comparison shown here is encouraging, further work is needed to understand how to quantify the relevant extent of plasticity that is predictive of failure. 

\end{document}